\documentclass[12pt]{article}

\pdfoutput=1
\usepackage{amsfonts,amssymb}
\usepackage{amsmath}
\usepackage{epsfig, graphicx}

\makeatletter\@addtoreset{equation}{section}\makeatother

\newcommand{\bea}{\begin{eqnarray}}
\newcommand{\eea}{\end{eqnarray}}
\newcommand{\be}{\begin{equation}}
\newcommand{\ee}{\end{equation}}

\def\T11{{T}^{1,1}}
\def\bear{\begin{eqnarray}}
\def\eear{\end{eqnarray}}

\newcommand{\pa}{\partial}
\newcommand{\tr}{{\rm tr}}
\newcommand{\comment}[1]{}

\newcommand{\pasl}{\pa\kern-.55em /}

\newcommand{\ksl}{k\kern-.55em /}

\newcommand{\vev}[1]{\langle #1\rangle}

\DeclareFixedFont{\xiiss}{OT1}{cmss}{m}{n}{12}
\DeclareFixedFont{\ixss}{OT1}{cmss}{m}{n}{9}
\DeclareFixedFont{\cmrnine}{OT1}{cmr}{m}{n}{9}
\newcommand{\field}[1]{\mathbb{#1}}

\newcommand{\BR}{{\field R}}

\newcommand{\CCs}{\hbox{\ixss C\kern-.4emI}}
\newcommand{\ZZs}{\hbox{\ixss Z\kern-.4emZ}}

\renewcommand{\em}{\it}
\renewcommand{\title}[1]{\vbox{\center\LARGE
\bf\mathversion{bold}{#1}\mathversion{normal}
}\vspace{5mm}}
\renewcommand{\author}[1]{\vbox{\center#1}\vspace{5mm}}
\newcommand{\address}[1]{\vbox{\center\em #1}}
\newcommand{\email}[1]{\vbox{\center\tt#1}\vspace{5mm}}

\begin{document}

\begin{titlepage}
\begin{center}
\vspace{5mm}
\hfill {\tt MAD-TH-10-07}\\
\vspace{30mm} 
{\font\titlerm=cmr10 scaled\magstep4 {\titlerm 
Dynamical tachyons on fuzzy spheres}}

\bigskip

\author{\large David Berenstein$^{1,2}$ and Diego Trancanelli$^{1,3}$}
\address{$^1$Department of Physics, University of California, Santa Barbara, CA 93106 \\
\medskip
$^2$Institute for Advanced Study, School of Natural Science, Princeton, NJ 08540\\
\medskip
$^3$Department of Physics, University of Wisconsin, Madison, WI 53706\\}

\email{dberens@physics.ucsb.edu,\quad
dtrancan@physics.wisc.edu}

\end{center}

\abstract{
\noindent
We study the spectrum of off-diagonal fluctuations between displaced fuzzy spheres in the BMN plane wave matrix model. The displacement is along the plane of the fuzzy spheres.  We find that when two fuzzy spheres intersect at angles classical tachyons develop and that the spectrum of these modes can be computed analytically. These tachyons can be related to the familiar Nielsen-Olesen instabilities in Yang-Mills theory on a  constant magnetic background. Many features of the problem become more apparent when we compare with maximally supersymmetric Yang-Mills on a sphere, of which this system is a truncation.  We also set up a simple oscillatory  trajectory on the displacement between the fuzzy spheres and study the dynamics of the modes as they become tachyonic for part of the oscillations. We speculate on their role regarding the possible thermalization of the system. 
}

\vfill

\end{titlepage}

\tableofcontents


\section{Introduction}

In gravitational theories it is expected that very high energy scattering at  small impact parameter is dominated by the formation and evaporation of black holes. Our basic understanding of how this might happen in a semiclassical approximation has led to Hawking's paradox, the statement that black holes might lose information with the scattering being non-unitary \cite{Hawking}. The same understanding that has led to this paradox shows that black holes have a very large entropy and that they also have a temperature. Both of these features cannot be explained by classical means.

The gauge/gravity duality has given us a way, in principle, to formulate these phenomena in a unitary framework, where the black hole intermediate state should be described by some approximately thermal object with large entropy in a dual quantum field theory.  The details of such approximately thermal field theory configuration, however, have not been worked out so far. Large black holes in $AdS$ have positive specific heat \cite{HawkingPage} and can be readily identified with a thermal state in the field theory~\cite{Witten:1998qj}. On the other hand, small black holes are not stable objects (since they evaporate) and a complete description of the system would require a theory of how such small black holes form and evaporate. An initial state that produces the black hole is not thermal, but the dynamics should be such that the system thermalizes rapidly (the black hole formation and ringing suggests this kind of picture \cite{SS}). The ensuing evaporation of the black hole shows that the system should be thought of as being out of equilibrium, but very long lived. For this second stage of evaporation one would need a detailed analysis of how the degrees of freedom -- the entropy -- of the black hole escape from it, if one is to claim to have really solved the black hole information paradox. 

An attempt to describe the initial stages of thermalization of small black holes\footnote{These are black holes which are much smaller than the radius of the $AdS$ geometry and behave as ten-dimensional Schwarzschild black holes \cite{Horowitz:2000kx}.} was done for ${\cal N}=4 $ super Yang-Mills theory in \cite{AsplundB}, but the details of the dynamics in that context are still poorly understood. In general,  problems of thermalization in full-fledged quantum field theories are hard to approach, especially at strong coupling. A promising strategy to try to address these issues is therefore to look for examples with fewer degrees of freedom. This paper is a first step in this direction, where we show that formulating the problem of black hole thermalization in a simpler setting has various advantages.\footnote{Other studies of thermalization based on toy models have been carried out  in \cite{IP}, but part of the system there already begins in a thermal state and the process of black hole production is not addressed.} One of such settings is represented by matrix quantum mechanical systems.

 The BFSS matrix model  \cite{BFSS} is the prime example of a matrix quantum mechanics. It is dual to M-theory on a discrete light-cone quantization of flat space and it has been argued that it accommodates black holes, which have been analyzed in various works \cite{BFKS}. The geometry of the black hole horizon can be related to properties of the matrix quantum mechanics and to the presence of tachyons  \cite{KLif}. Some of the studies of the BFSS model have also included numerical analysis based on lattice methods in Euclidean field theory \cite{Latt}. However, a detailed analysis of the formation and thermalization properties of these matrix black holes is still missing. Moreover, a numerical instability arises in these Euclidean thermal systems, exactly because eigenvalues can leave the black hole state. As such, the Euclidean ensemble does not exist, as one would expect when studying systems that cannot be in true equilibrium.

This problem can be fixed by considering instead the BMN matrix model \cite{BMN}, where the large volume instability of having a moduli space of vacua with a continuous spectrum is cured and the simulations make sense \cite{Catterall:2010gf}. The BMN matrix model is in fact a mass deformation of the BFSS model which lifts the flat directions and has no problems from this point of view. As the BFSS model, it has an M-theory interpretation, as it describes the discrete light-cone quantization of the theory on a plane wave (rather than on flat space). Another nice feature of the BMN matrix model is that it can be thought of as an $SU(2)$ invariant sector of ${\cal N}=4$ super Yang-Mills theory on $S^3 \times \mathbb{R}$ \cite{KKP}. This means that certain classes of analysis of this model might also be fruitful to better understand the physics of that four-dimensional theory. 

Our main interest in this paper is to explore some time dependent dynamics, in a matrix quantum mechanics setting, that might help us understand how the initial stages  of black hole thermalization might proceed. For this, we need some simple configurations with given initial conditions that are similar to the problem of scattering gravitons at high energies. We can then study more precisely the dynamics of the modes that produce the thermalization. In the analysis of \cite{AsplundB}, it was argued that particle production in off-diagonal modes connecting eigenvalues was responsible for generating the entropy of the black hole configuration and for trapping the eigenvalues. If we try to mimic this intuition, we would expect it to work very similarly for the BFSS matrix model: we could think of scattering two graviton states with large matrices and to hopefully see the black hole form. Unfortunately, we are unable to do this because these graviton states in the BFSS matrix model are very poorly understood: they are bound states at threshold and  therefore we would need a detailed understanding of their wave function, which we do not have, to do the analysis. 

For this reason and for the reasons mentioned above, we turn instead our attention to the BMN matrix model. Here the set of ground states is richer than in the BFSS model and instead of scattering eigenvalues off each other we can scatter more complicated ground state configurations. These configurations are given by sets of concentric fuzzy spheres. Each such  fuzzy sphere is a giant graviton  \cite{GG} that can grow in size due to the Myers effect \cite{Myers:1999ps}.
For each fuzzy sphere, one has a decoupled $U(1)$ set of degrees of freedom that correspond to the center of mass motion of the corresponding membrane object. These degrees of freedom can be excited leaving the rest of the $SU(N)$ degrees of freedom unaffected. This means that the fuzzy spheres can be rigidly moved around without being deformed and their geometry is simple to analyze. Thus, it is a simple matter to find reasonable initial conditions for these fuzzy spheres. By aiming them properly, we can make them collide (namely, have two such fuzzy spheres intersect each other at some time $t$). The dynamics near the intersection is interesting. From the point of view of string theory, these are configurations with branes at angles, so there is a possibility of having tachyons form at these intersections \cite{Berkooz:1996km}. Moreover, even in the absence of tachyons, modes near the intersections are light. As we move the branes past each other, these modes have a time dependent mass and in general we expect copious particle production in these modes if the adiabatic approximation breaks down. At weak coupling this is similar to the problem of preheating in inflation \cite{Kofman:1994rk}. In our case, the matrix model black hole would  result from thermalization after moduli trapping \cite{Kofman:2004yc}.\footnote{The trapping between D0-branes was analyzed in \cite{DKPS}.} A further advantage that we have in this setup over the BFSS model is that the classical solutions corresponding to branes crossing each other are periodic: the branes can cross repeatedly and the fluctuations that are produced in this way have time to grow through many repetitions until they cause a large back-reaction.

The main goal of this paper is to describe in detail the mass spectrum of the bosonic modes stretching between fuzzy spheres, as we displace the spheres along a direction longitudinal to their world-volume and we make them intersect. Our calculation is a generalization of the analysis done in \cite{VRetal} to study the fluctuations around concentric fuzzy spheres.\footnote{Some studies of the thermodynamics of these configurations include the works in \cite{fuzzytherm}. Other important result about protected multiplets can be found in \cite{BMNextra}. } While fluctuations around some fuzzy sphere configurations have already been considered in the past by other groups, in this paper  we focus our attention on a different and novel set of configurations. In previous studies, the displacements of the fuzzy spheres are in the transverse directions to the spheres and the distance between the spheres (or other branes \cite{Bak:2002rq}) does not change in time, or does so in such a way that the time dependence of the off-diagonal modes is perturbatively small \cite{fuzzycalc}.

We will argue that in our setup there is a spectrum of tachyons at the intersection locus of the fuzzy spheres. The intersection locus being one-dimensional gives us a one-dimensional tower of such tachyons. There is such a tachyon even for two D0-branes at some finite distance. These tachyons can have large negative mass when at least one of the fuzzy spheres corresponds to a large matrix,  so that any small fluctuation in these modes can grow rapidly and might lead to fast thermalization. A detailed analysis of the evolution of the system following the tachyon production is beyond the scope of the present work, but we sketch nonetheless a picture of what happens in a simple case.

The paper is organized as follows. In Section~\ref{sec-2} we review briefly the BMN plane wave matrix model and the classical vacua of the theory. We present a novel derivation of the spectrum of fluctuations around concentric fuzzy spheres making use of the aforementioned relation between the BMN matrix model and ${\cal N}=4$ super Yang-Mills. In Section~\ref{sec-3} we consider a classical solution of the BMN matrix model consisting of two fuzzy sphere which are not concentric, but displaced along a direction longitudinal to their world-volume. After giving a geometrical characterization of the fluctuations around this configuration, we proceed in Section~\ref{sec-4} with a detailed computation of the spectrum of off-diagonal fluctuations, both along the transverse directions to the spheres and along the longitudinal ones. We find that certain longitudinal modes can become tachyonic. In Section~\ref{sec-5} we study the time dependence of these tachyonic modes as we allow the displaced fuzzy spheres to oscillate one towards the other along the direction of the displacement. Some concluding remarks can be found in Section~\ref{sec-6}, while in the Appendix we present an alternative derivation of the spectrum of longitudinal fluctuations.


\section{The BMN matrix model }
\label{sec-2}

The BMN matrix model \cite{BMN} is a massive deformation of the BFSS matrix model \cite{BFSS}. The latter is obtained from the dimensional reduction of ten-dimensional ${\cal N}=1$ super Yang-Mills down to $0+1$ dimensions and has an action given by
\begin{equation}
S_{BFSS}= \frac 1{2 g^2}\int dt \,  \tr \left( (D_t X^I)^2+ \frac 1{2} [X^I,X^J]^2\right) +\hbox{fermions}\,,
\label{S_BFSS}
\end{equation}
where $X^I$ ($I=1,\ldots,9$) are nine hermitian matrices. The covariant time derivative is given by
\begin{equation}
D_t X^I = \partial_t X^I-i [A_t, X^I]
\end{equation}
and $g$ is a dimensionful coupling constant that can be removed by rescaling the fields and the time coordinate. It can be set to one, if desired, or factored out of the action and interpreted as determining $\hbar$. We will not work in detail with the fermions in this paper, so we shall just suppress them from now on.
The Hamiltonian of this system (in the $A_t=0$ gauge) is given by
\begin{equation}
{\cal H} = \frac{1}{2}\tr\left(g^2 (\Pi^I)^2- \frac 1{2g^2} [X^I,X^J]^2\right)\,.
\end{equation}

The BMN matrix model system is a massive deformation of (\ref{S_BFSS}) that preserves all 32 supersymmetries. It also preserves a diagonal set of modes that decouple and constitute a system of free degrees of freedom. These are the `center of mass motion' degrees of freedom in the BFSS matrix model.
The BMN matrix model splits the $X^I$ into two groups of variables: $X^{1,2,3}$, that we will label $X^i$, and $X^{4,\dots, 9}$, that we will label $Y^a$. The action includes additional terms given by
\begin{equation}
S_{BMN} = S_{BFSS} - \frac{1}{2g^2}\int dt\, \tr\left(\mu^2 (X^i)^2+\frac{\mu^2}{4}(Y^a)^2
+2\mu i \,\epsilon_{\ell jk}X^\ell X^jX^k \right) 
\,.
\label{BMNaction}
\end{equation}
In the conventions above, $\mu$ is real and has been rescaled by a factor of $3$ with respect to $\cite{BMN}$. It has units of frequency, as $X^I$.
The equations of motion following from this action are
\begin{eqnarray}
\ddot X^i &=& -\mu^2X^i -3 i\mu\,\epsilon^{ijk}X^jX^k-\left[\left[X^i,X^I\right],X^I\right]\,,\cr
\ddot Y^a &=& -\frac{\mu^2}{4}Y^a -\left[\left[Y^a,X^I\right],X^I\right]\,.
\label{EoM}
\end{eqnarray}

It is convenient for our study to recast the potential for the $X^i$ fields in the following form
\begin{equation}
V^{(X)}_{BMN}= \frac 1{2g^2}\tr  \left[\left(i [X^2,X^3]+\mu X^1 \right) ^2
+\left( i[X^3,X^1]+\mu X^2 \right) ^2+\left(i [X^1,X^2]+\mu X^3 \right) ^2\right]\,.\label{eq:VXBMN}
\end{equation}

In this paper we will choose to rescale $\mu \to 1$ by both scaling the matrices and the time coordinate. This can always be done. The overall action then has a factor of $1/g^2$ in front. This serves as a calibration of $\hbar$: we are free to absorb $g^2$ in the definition of $\hbar$. This has no effect on the classical physics except for the global normalization of the energy units. It is only when we quantize the theory that the value of $g$ will be important and it will characterize the strength of quantum effects or fluctuations around classical solutions of the dynamics.

An important point to notice is that the potential  is a sum of squares of hermitian matrices. The $V_{BMN}^{(X)}$  is positive definite and the same is true for the other terms in the potential: the quadratic terms in $Y^a$ are obviously a sum of squares and the rest of the terms in (\ref{S_BFSS}) are commutator squared terms with the right sign to make them positive definite.

Another very useful result to recall is that the BMN matrix model can be obtained by considering a truncation to the $SU(2)$ invariant configurations of ${\cal N}=4 $ super Yang-Mills on $S^3\times \BR$ \cite{KKP}. The sphere has a $SO(4)\simeq SU(2)_L\times SU(2)_R$ symmetry and the $SU(2)_L$ invariant sector of the theory gives exactly the model above. In that case, $g$ is identified with the coupling constant of ${\cal N}=4$ super Yang-Mills. The $X^i$ degrees of freedom  arise from the gauge connection on $S^3$ and in the usual field theory analysis would have mass 2. The $Y^a$ degrees of freedom arise instead from the scalars $\phi^a$ of the super Yang-Mills theory and would ordinarily have mass equal to one. This corresponds to setting $\mu=2$ in the BMN model above. Equivalently, we can think of this as quantizing the ${\cal N}= 4$ theory on a sphere of radius equal to 2, rather than one. This $SU(2)_L$ invariant reduction is a convenient device to calculate sometimes and it also explains why the potential is a sum of squares.


\subsection{Fuzzy sphere ground states and fluctuations}
\label{sec-grouptheory}

The ground states of the BMN matrix model are those that have zero energy. These are characterized by $Y^a=0$ and also by
\begin{equation}
[X^i, X^j]= i \epsilon_{ijk} X^k\,,
\end{equation}
where we have set $\mu=1$. These are obtained by requiring that each of the individual squares in the potential vanish.

The most general solution to this equation is given by a possibly reducible lie algebra representation of $SU(2)$, where $X^i \simeq \bigoplus_\alpha L_{(n_\alpha)}^i$, with the sum indicating a sum over irreducible representations of $SU(2)$. Each irreducible representation (which we can label by the size of the representation $n=2j+1$, or by the maximal spin $j$) gives rise to a fuzzy sphere configuration. In the BMN model these are interpreted as giant graviton membranes of the plane wave limit M-theory dual. 

If we fix the total size of the matrices $N$,  we can set $N=\sum_\alpha n_\alpha$. The complete set of vacua of the theory is characterized by these possible splittings. This is equal to the partitions of $N$. 

Given such a configuration, we can think of the ground state vacuum expectation values as being block diagonal. We can then ask what is the spectrum of off-diagonal fluctuations connecting two of these fuzzy spheres. This has been analyzed in detail in \cite{VRetal}.

Here, we will reproduce that result using a slightly different calculation. The idea is to remember that this matrix quantum mechanics is an $SU(2)_L$ invariant reduction of the $U(N)$ ${\cal N}=4$ super Yang-Mills theory on $S^3\times \BR$. All of the fuzzy sphere vacua are related to each other by a gauge transformation in the theory on $S^3$ \cite{LinMalda}. The gauge transformation that relates them is not $SU(2)_L$ invariant, so the modes that are considered  $SU(2)_L$ invariant get shuffled.
The gauge transformation uses the obvious map $S^3\to SU(2)$, since they are identical spaces.
This $SU(2)$ has to be embedded in the gauge group. 

For a fuzzy sphere configuration $N=\sum_\alpha n_\alpha$, we embed $SU(2)$ into $U(N)$ by the action on the fundamental of $U(N)$ according to $\bigoplus_\alpha R_\alpha= \bigoplus_\alpha (n_\alpha)$, where we are labeling the representations by $U(N)$. This $SU(2)$ embedding can also be thought of as $SU(2)_L$: the translation on the sphere by $SU(2)_L$ generates the sphere itself. Also notice that the fuzzy sphere configurations are $SU(2)_L$ invariant only if an $SU(2)_L$ rotation is accompanied by a compensating gauge transformation. 

Now let us do some group theory. The Fock space spectrum of physical polarizations of fluctuations of the $A=0$ vacuum is given by
\begin{eqnarray}
{Spec}( \phi) &=& \bigoplus_j (j,j)\,,\cr
{Spec} (A) &=& \bigoplus_j (j,j+1)\oplus(j+1,j)\,,
\end{eqnarray}
where we are using the spin notation for the representations. The first line indicates the spectrum of representations of the scalar fluctuations, while the one of the second line represents the transverse fluctuations of gluons. The energy for the scalars $\phi$ is $2j+1$, while for the vectors it is $2j+2$.

When we consider the embedding on matrices, we have that the off diagonal block connecting block $n_1$ and $n_2$ transforms as $j_1 \otimes j_2$ with respect to $SU(2)$. This is, the matrices transform as $\bigoplus_{j'=|j_1-j_2|}^{j_1+j_2}  j'$ with respect to $SU(2)_L$. If we tensor these together, we find that for the scalars we need to take the tensor product
\begin{equation}
\bigoplus_{\tilde j} \tilde j \otimes (j,j) \simeq \bigoplus_{\tilde j} \bigoplus_{ j'= |\tilde j-j|}^{\tilde j +j} ( j',j)\,.
\end{equation}
We can only have $SU(2)_L$ singlets if $j'=0$, so we find that the singlet sector is given by the case where $j=\tilde j$. We find this way that 
\begin{equation}
\bigoplus_{\tilde j} \tilde j \otimes (j,j)\big|_{singlet} \simeq \bigoplus_{\tilde j} ( 0 ,\tilde j)\,,
\end{equation}
with energy $2\tilde j+1$. Notice that $\tilde j$ is either only half integer or only integer, as per the usual rules of angular momentum. 

Similarly, we find that for the vectors
\begin{equation}
\bigoplus_{\tilde j}\tilde j \otimes \left((j,j+1)\oplus(j+1,j)\right) \big|_{singlet} \simeq \bigoplus_{\tilde j} (0, \tilde j+1)\oplus (0, \tilde j-1)\,,
\end{equation}
with energies $2\tilde j+2$ and $2\tilde j$ respectively. For the special case $\tilde j=0$, the representation $(0,-1)$ is not counted.

Properly normalizing to the value of $\mu$ we have used ({\it i.e.}, dividing by 2 the energies above), we get that on the fuzzy sphere the scalars $Y^a$ have a spectrum under $SU(2)_R$ rotations given by
\begin{equation}
Spec(Y)= \bigoplus_{\tilde j} \tilde j\,,
\end{equation}
with energies $\tilde j +1/2$, where $\tilde j $ is only integer or half integer and it runs between 
$|j_1-j_2|$ and $j_1+j_2$ in integer steps.

Similarly, for the $X^i$ fields, we get a group decomposition
\begin{equation}
Spec(X) = \bigoplus_{\tilde j} (\tilde j+1)\oplus (\tilde j-1)\,,
\end{equation}
with energies $\tilde j+1$ and $ \tilde j$ respectively. Remember that $\tilde j$ is always integer or half integer.
The same analysis can be done for fermions, we will not do this here. 

The representations $\tilde j$ appearing in the decomposition of the off-diagonal matrix fluctuations of $Y^a$ are called fuzzy monopole harmonics. They are fuzzy spherical harmonics if the two representations have the same dimension.

For the $X^i$ variables, the $\tilde j$ representations appearing in the decomposition are called fuzzy monopole vector (tensor)  harmonics and fuzzy vector (tensor) harmonics if the two representations have the same size.

Notice that we only described the physical fluctuations. The typical configurations also have zero modes due to gauge transformations. These zero modes have been projected out. If we add them, we get that $X$ has an additional set of zero modes that are described by the fuzzy monopole spherical harmonics. There is one more thing that needs to be remarked: the zero mode associated to rotations in the diagonal $U(1)$ is absent as nothing is charged under it.

For matrices of the same size the theory has an enhanced $SU(2)$ unbroken
gauge symmetry and the modes organize themselves into triplets and singlets of $SU(2)$. This generalizes to when one has more coincident fuzzy spheres.


\section{Kicking the spheres}
\label{sec-3}

We now consider the following classical solutions of the BMN matrix model:\footnote{These configurations are non-BPS.}
\begin{eqnarray}
\vev{X^i}= \begin{pmatrix} L_{(n_1)}^i +\Re e(b^i_1*\mathsf{1}_{(n_1)} \exp(i t))& 0 &\dots\\
0&L_{(n_2)}^i+\Re e(b^i_2*\mathsf{1}_{(n_2)} \exp(i t))&\dots\\
\vdots&\vdots &\ddots 
\end{pmatrix}\,.
\label{vev}
\end{eqnarray}
We have turned on modes proportional to the identity that decouple on each diagonal block and describe the center of mass motion of the fuzzy spheres. These modes leave the shape of the spheres invariant, but move their centers of mass in time. A similar kicking of the spheres can be done by turning on a displacement $b^a_\alpha$ along the $Y^a$ directions
\begin{eqnarray}
\vev{Y^a}= \begin{pmatrix} \Re e(b^a_1*\mathsf{1}_{(n_1)} \exp(i t/2))& 0 &\dots\\
0&\Re e(b^a_2*\mathsf{1}_{(n_2)} \exp(i t/2))&\dots\\
\vdots&\vdots &\ddots 
\end{pmatrix}\,.
\end{eqnarray}
Notice that in order to obey the equations of motion (\ref{EoM}) the $Y^a$ directions must have a frequency that is half the frequency of the $X^i$ directions. For the rest of this paper we shall set $\vev{Y^a}=0$.

We restrict now our analysis to the case of two diagonal blocks, with ranks $n_1$ and $n_2$. We will first study the dynamics of the off-diagonal modes connecting the two fuzzy spheres when they are large, {\it i.e.} with $n_1$, $n_2,$ and $ n_1-n_2$ being large.   For the configurations with $b^i_\alpha=0$, the off-diagonal modes have a high angular frequency starting at $n_1-n_2$, so this frequency is much larger than the frequency of the zero mode that we are kicking. This means that small fluctuations on these degrees of freedom are expected to be adiabatic for a typical $b^i_\alpha$. 

There is a clear procedure to analyze this type of configurations. One thinks of this as a Born-Oppenheimer approximation where one can solve the dynamics of fast degrees of freedom as we freeze the slow degrees of freedom. The fast degrees of freedom are the off-diagonal modes and the slow degrees of freedom are going to be the zero modes that we turned on. 
They are only fast relative to the motion we have described if their spin is large.
We will analyze these  by  freezing the result at some time $t$ and studying the spectrum of quadratic fluctuations of the off-diagonal modes for that frozen configuration. Since $t$ is fixed for the analysis, we can use the rotational symmetry of the system to align $\vec b_{1}- \vec b_{2}$ along the $3$-axis ({\it i.e.} only $b^3_1-b^3_2\neq 0$). Moreover, a combination of the form $n_1 \vec b_1+n_2 \vec b_2$ is proportional to the center of mass of the whole system and is a decoupled degree of freedom. This means that all the dynamics we are interested in depends only on $\vec b_{1}-\vec b_2$. We can without loss of generality set $\vec b_1=0$. Also, we can choose the displacements to be real.

In the rest of this section we will give a geometric characterization of what type of results we expect to find. A detailed analysis of the spectrum of these fluctuations will then follow in the next section. 


\subsection{Geometric characterization of the dynamics}

These matrix configurations can be described geometrically in the M-theory plane wave geometry. A fuzzy sphere of rank $n$ corresponds to an M2-brane giant graviton of size proportional to its light-cone momentum $n$ (the proportionality constant depends on conventions). In matrix coordinate units, $n$ acts as a cut-off on the range of the angular momentum (see \cite{Madore}), so that the size of the sphere is also proportional to the maximum eigenvalue $j$.  Defining the radius of the sphere as given by the distance to the center of mass (this is also done in Matrix Theory \cite{Kabat:1997im}), we get
\begin{equation}
R^2 = (X^1- x^1_{cm})^2+(X^2-x^2_{cm})^2+(X^3-x^3_{cm})^2= j(j+1)\,.
\end{equation}
At large $j$ we have that the radius is $R= j+1/2+ o(1/j)$. Notice that the spacing of the radii between different $j$ is essentially $\Delta j$. These M2-branes are described as D2-branes. Off-diagonal modes connecting different such configurations are interpreted as strings stretching between the D2-branes, while the eigenvalues are interpreted as D0-branes.
It is standard to think of the D2-branes as branes that have absorbed D0-branes and as a consequence have a strong magnetic field on their world-volume \cite{Dougl}. As a matrix of rank $n$ has $n$ eigenvalues, this is the magnetic flux threading the D2-brane sphere. A string ending on the D2-brane is charged under this magnetic field and experiences a magnetic  monopole flux of strength $n$.  As is well known, if we consider a charged scalar degree of freedom on a sphere with a magnetic monopole background and if we restrict to the lowest Landau level, the wave functions carry angular momentum and can be argued to be localized on the sphere so that they cover an area of order $1/n$. The angular momentum points in the direction on which we localize the wave function on the sphere. If we reverse the charge, the angular momentum points in the opposite direction.

If we have two fuzzy spheres of sizes $j, j'$ at rest they can be described by two concentric spheres of ranks $n, n'$.  Now let us consider strings stretching between  them. If we attach $n$ states to one sphere and $n'$ states to the other sphere, we get a total of $nn'$ possible strings. These strings will have an endpoint on a sphere that is associated to a  positive charge and the other end on the other sphere will have the opposite charge (the string theory is oriented). The effective magnetic flux that the particle sees is $n-n'$. This describes the minimal angular momentum that the modes connecting the fuzzy spheres can have: $|j-j'|$. This is also the length of a string stretched from the north pole of one fuzzy sphere to the north pole of the other one. 

\begin{figure}[ht]
\begin{center}
\includegraphics{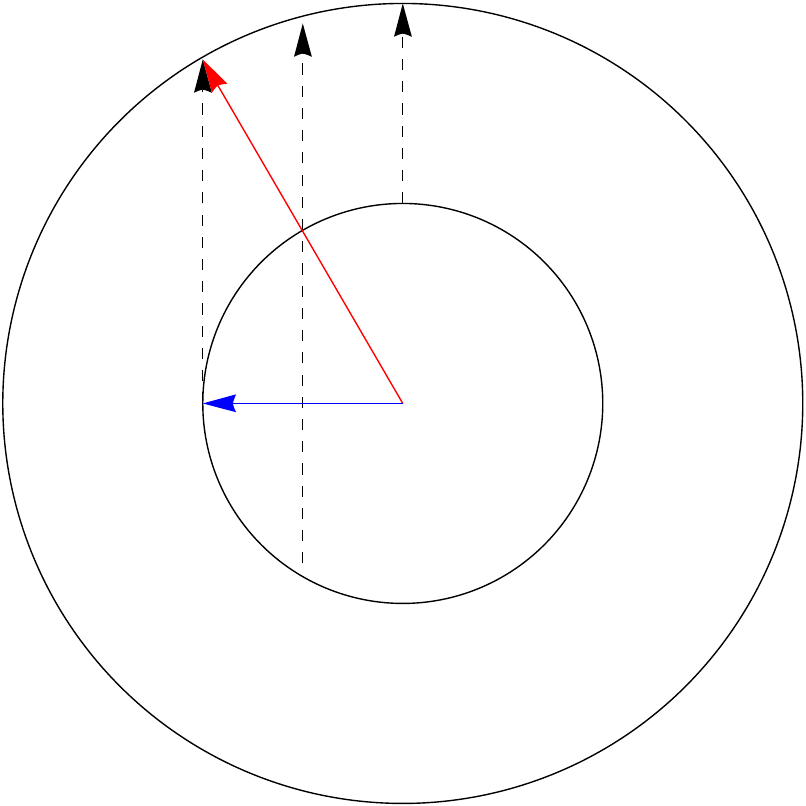}
\caption{An off-diagonal mode can be thought of as a string stretching between two D2-branes. In this figure we consider the case of zero displacement, {\it i.e.} concentric spheres. The angular momentum of a state is given by subtracting the position vectors of the string endpoints. Using the symmetries of the system, the angular momentum vector can always be chosen to be aligned along the vertical axis. 
\label{fig:ang}}
\end{center}
\end{figure}

The string of maximum length has angular momentum given by $j+j'$ and a mass of order $j+j'$ in appropriate units. The angular momentum vector can be obtained by taking the difference of the position vectors  of the endpoints of the string. This is depicted in figure \ref{fig:ang}. The angular momentum vector of the string state points parallel to the string. 
Also, the geometric length is proportional to the length of the angular momentum vector (this also happens for Non Commutative field theories on the Moyal Plane \cite{BigattiSusskind}). We depict in the figure various highest weight states. The longest string goes from the north pole of one fuzzy sphere to the south pole of the other. Notice also that each string endpoint can be thought of as occupying a uniform fixed area on each sphere. The sphere with $n_1$ eigenvalues has $n_1$ such patches and similarly the second one, with $n_2$ eigenvalues, will be divided into $n_2$ patches. Each of these is to be though of as a D0-brane end-point (region) on the sphere.

If we compare with the spectrum of the $Y^a$ fluctuations, we get a precise matching between the possible values of angular momentum we compute geometrically in this way and the ones we obtain from the field theory calculation. These are transverse polarizations of the strings to the three directions in which the branes are embedded. For polarizations of the string modes in the brane 3-plane, they have extra spin: indeed, they carry one unit of spin that is either along the direction of the string, or opposite to it (only transverse polarizations appear on the string) and one can match this to the values of angular momentum of the $X^i$ fluctuations as well. Again, the geometric estimate of the mass is good enough. Notice that there are in general correction of order one to the mass. 

\begin{figure}[ht]
\begin{center}
\includegraphics[scale=0.7]{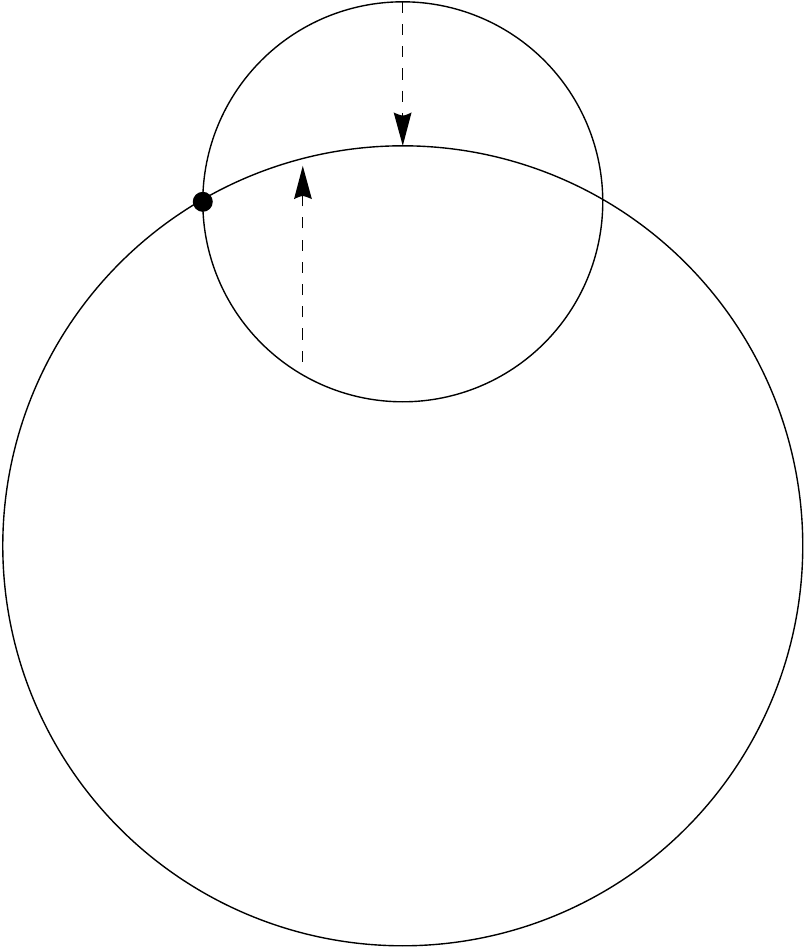}
\caption{The length of the strings changes as we displace the fuzzy spheres. 
\label{fig:displaced}}
\end{center}
\end{figure}

Now we can consider what happens when we displace the spheres (see figure \ref{fig:displaced}). It is clear that the density of the string endpoints on each fuzzy sphere is not going to change. This is because this is roughly the density of eigenvalues per unit area. Moreover, since the end points of the string are charges in a very strong magnetic field, the magnetic field is good at keeping them from changing positions. Roughly, we should imagine that the string ends are stuck to a particular D0 brane and that these D0 branes are heavy and don't get moved around much by the force of a string. Indeed, in perturbation theory the tension of the D-branes is large  \cite{Dai:1989ua}. So we can argue that the endpoint location of the string on each fuzzy sphere will not change, nor how we think about its angular momentum in the $3$-direction (the one that is preserved by the configuration).  However, the length of the string will change. 

If we use the same labels for the string endpoints as before, that is we label them by their angular momentum, we find that the length is given by 
\begin{equation}
L ^2\simeq  (\Delta L^1)^2+(\Delta L^2)^2+(\Delta L^3-b)^2\,,
\end{equation}
where $b$ is the displacement. The masses should then be roughly given by
\begin{equation}
M^2\simeq  (\Delta \vec L)^2 - 2 b \Delta L^3 + b^2\,. 
\label{eq:geommass}
\end{equation}
The mass formula will attain the minimum value on a sphere for fixed $(\Delta \vec L)^2$ when $\Delta L^3$ takes either the maximum or the minimum value, depending on the sign of $b$.

Clearly, this value will be minimized when the spheres touch. The states with minimum energy have their spin aligned along the $3$-axis shown in the figure (the strings of length zero do not have an horizontal component of $\vec L$). There will be corrections to this simple geometric formula. This can be seen  from comparing the values of the energy when the spheres are concentric to those that are obtained from the exact answer. These corrections to the mass squared are of order $j$ and can in principle make some of the modes tachyonic. This is what one can expect from the intuition of branes at angles  \cite{Berkooz:1996km}. To check this requires doing a computation, that we do in the next section. What is important to notice is that if there are tachyons, they are localized near where the spheres touch. In this examples this is a circle, so one can expect that these degrees of freedom are like a one-dimensional field theory set of modes. 

Notice that the naive geometrical correction is proportional to $\Delta L^3$. If we treat it as an operator in the theory of angular momentum it commutes with $(\Delta \vec L)^2$, so even though the configurations break the rotational symmetry, the spherical harmonic representation of the states should remain diagonal. This is similar to what happens in  the computation of the Zeeman splitting in the hydrogen atom. We will see this explicitly when we do the full computation. Also notice that the most tachyonic mode will depend on the displacement, because of all the vertical strings the one that is the shortest depends on the precise value of $b$. This means that since in our kicked sphere solution the displacement is changing with time, which off-diagonal mode is the most tachyonic depends on time. If all of  these start condensing, one expects that, since they carry angular momentum, the axial symmetry will be broken classically: the spheres will deform non-uniformly and change shape.  The $L^3$ will remain a  constant of motion and the axial symmetry will only be restored quantum mechanically by averaging over the orientation of the resulting shape. The axial symmetry can be restored later again classically if the system thermalizes and any coarse grained observation becomes sufficiently homogeneous (this usually requires large $N$). Such effects would be expected when we form a non-rotating black hole in the gravity theory.


\section{Spectrum of fluctuations}
\label{sec-4}

Now we are ready to start calculating the spectrum of fluctuations of off-diagonal modes for displaced fuzzy spheres. As we have described above, at large $N$ the off-diagonal modes connecting two fuzzy spheres are generically heavy, except where the fuzzy spheres intersect. We have also argued that geometrically it seems that the basis of spherical harmonics is preserved.

We will expand both the $X^i$ and $Y^a$ variables in spherical monopole harmonics. This is because the off-diagonal matrices have the same rank. The fuzzy vector spherical harmonics will be particular linear combinations of these. The basis of matrices will be labeled by $Y_{\ell m} \in Hom(n_2, n_1)$, which explains how we think of these as matrices. The conjugate harmonic is 
$Y_{\ell m} ^\dagger = (-1)^m Y_{\ell, -m} \in Hom(n_1,n_2)$. They are normalized so that 
\begin{equation}
\tr( Y_{\ell m}^\dagger Y_{\ell' m'}) = \frac 12 \delta_{\ell\ell'}\delta_{mm'} \,.
\end{equation}

The matrices $Y^a$ that are off-diagonal and connecting the two fuzzy spheres will be hermitian and expanded as follows
\begin{equation}
Y^a = \sum_{\ell,m} y^a_{\ell m} Y_{\ell m}+ (y^a_{\ell m})^* Y_{\ell m}^\dagger\,,
\end{equation}
or, writing the blocks more explicitly,
\begin{equation}
Y^a = \sum_{\ell,m}\begin{pmatrix}
0 & y^a_{\ell m} Y_{\ell m}\\ (y^a_{\ell m})^* Y^\dagger_{\ell m} & 0
\end{pmatrix}\,.
\label{expYblocks}
\end{equation}
We will do the same for the $X^i$ matrices. The difference is that the $X^i$ matrices will also have a vacuum expectation value as in (\ref{vev}).
As argued above, we can take 
 \begin{equation}
 \vev{X^3}= \begin{pmatrix}
  L_{(n_1)}^3 &0\\
  0 & L_{(n_2)}^3+ b
  \end{pmatrix}\,,
 \end{equation}
where $b$ is the displacement of the second fuzzy sphere, while we take $\vev{X^1}$ and $ \vev{X^2}$ with no displacement. Our goal is to compute the masses of the fluctuations in terms of $\ell, m, n_1, n_2, b$ and any possible mixings between the states.  To do this, we will need the following commutation relations for the $Y_{\ell m}$
\begin{eqnarray}
{[L^3, Y_{\ell m}] } &=&   m Y_{\ell m}\,,\cr
{[L^+, Y_{\ell m}] }&=& \sqrt{(\ell -m)(\ell +m+1)} Y_{\ell m+1}\,,\cr
{[L^-, Y_{\ell m}]}  &=&\sqrt{(\ell +m)(\ell+1-m)} Y_{\ell m-1}\,,
\end{eqnarray}
and their adjoints
\begin{eqnarray}
{[L^3, Y^\dagger_{\ell m}] }&=& -mY_{\ell m}^\dagger\,,\cr
{[L^+, Y_{\ell m}^\dagger] }&=&-  \sqrt{(\ell +m)(\ell -m+1)} Y^\dagger_{\ell m-1}\,,\cr
{[L^-, Y_{\ell m}^\dagger] }&=&-  \sqrt{(\ell -m)(\ell +m+1)} Y^\dagger_{\ell m+1}\,,
\end{eqnarray}
where $L^\pm = L^1\pm i L^2$.

First we study the fluctuations of $Y^a$, which are easier to analyze, and then those of $X^i$. We will do the calculation for arbitrary values of $n_1, n_2$ with the configurations being off-shell (frozen at finite displacement).


\subsection{Transverse fluctuations}

To compute the spectrum of transverse fluctuations we need to expand the Lagrangian in fluctuations $\delta Y^a = \sum_{\ell, m} \delta  y^a_{\ell m} Y_{\ell m}+ (\delta y^a_{\ell m})^* Y_{\ell m}^\dagger$ keeping only the terms up to quadratic order. Since the $Y^a$ do not have a vev and they appear always at least quadratically in the action, we can just truncate the action to quadratic order in $Y^a$ and replace the $X^i$ by their vevs. There can be no mixing between $X^i$ and $Y^a$ (this is easiest to argue by the fact that the $Y^a$ carry $SO(6)$ symmetry labels and the configurations under study are $SO(6)$ invariant).
The expansion of the Lagrangian is straightforward and yields
\begin{eqnarray}
{\cal L}^{(Y)} &=& \frac12  \tr(\dot Y^a)^2-\frac18 \tr (Y^a)^2 + \frac 12 \tr [\vev{X^i},Y^a]^2\cr
&=&\frac 12 \tr(\dot Y^a)^2-\frac18 \tr (Y^a)^2  +\frac 12\tr[\vev{X^3},Y^a]^2+ \frac 12 \tr[\vev{X^+},Y^a][\vev{X^-},Y^a]\,.
\end{eqnarray}
The first thing we compute is the kinetic term, which is obviously given by
\begin{equation}
{\cal L}^{(Y)}_{kin} =\frac{1}{2} \sum_{\ell, m} |\delta \dot y^a_{\ell m}|^2\,.
\end{equation}
For $b=0$ it is easy to see that the potential can be written as \cite{VRetal}
\begin{equation}
{\cal L}^{(Y)}_{mass} = -\frac{1}{2} \tr \left[Y^a\left(\frac14 Y^a+\left[L^i,\left[L^i,Y^a\right]\right]\right)\right]\,.
\end{equation}
Since $[L^i,[L^i,Y_{\ell m}]]=\ell(\ell+1)Y_{\ell m}$, this implies that at $b=0$ we get that the mass terms are given by
\begin{equation}
{\cal L}^{(Y)}_{mass} = -\frac{1}{2}\sum_{\ell,m} \left(\ell+\frac 12\right)^2 |\delta y^a_{\ell m}|^2\,.
\end{equation} 
This matches of course with the result obtained in section \ref{sec-grouptheory} using group theoretical arguments. A more detailed description of these spherical harmonics can be found in \cite{Ishiki:2006yr}.

We now turn on $b$. Notice that schematically it is $\vev{X^3}= L^3 + b\begin{pmatrix}0&0\\
0&1\end{pmatrix}$, so that 
\begin{eqnarray}
[\vev{X^3}, \delta Y^a]&=& [L^3,\delta Y^a] + b \left[ \begin{pmatrix}0&0\\
0&1\end{pmatrix}, \delta Y^a\right]\cr
&=&  \sum_{\ell,m} (m -b) \delta y^a_{\ell m} Y_{\ell m}-(m-b)(\delta y^a_{\ell m})^* Y_{\ell m}^\dagger \,.
\end{eqnarray}
Notice that the addition of $b$ does not mix the $Y_{\ell m}$ with each other in the commutator and all it does is to replace $m\to m-b$ in the commutation relations.  This is what makes the computation so simple. This is what we were arguing for geometrically in the previous section.
When we square the expression above, we get the same result as when $b=0$, plus additional terms that are cross terms and a quadratic term in $b$
\begin{equation}
\omega^2_{\ell m} = \left(\ell+\frac12\right)^2 - m^2 + (m-b)^2 \geq 0\,.
\end{equation}
Notice that these are all positive, because $\ell \geq m$. The minimum possible value, fixing $\ell$ and $m $ but varying $b$, is given by $b=m$, in which case the frequency squared is
\begin{equation}
\omega_{\ell m }^2 = \left(\ell+\frac12\right)^2 - m^2
\end{equation}
and the minimum value this can acquire is given by $m=\pm\ell$, so that the mass is 
\begin{equation}
\omega_{\ell \ell }^2= \ell +\frac 14\,.
\end{equation}
 So, as we go to higher and higher $\ell$, we find that the mode is more and more massive at the place where it is lightest (namely for $b=m$). However this only grows as $\sqrt \ell$, which is subleading to the typical value of the frequency which is of order $\ell$. If we compare with our geometric result in equation \eqref{eq:geommass} we find that it matches it very closely and it is exactly the same if we interpret $(\Delta \vec L)^2$ with the usual quantum value $\ell(\ell+1)$ plus the 1/4 from the background curvature of the plane wave.


\subsection{Longitudinal fluctuations}
 
Now we analyze the fluctuations of the $X^i$ fields. This is trickier than for the $Y^a$ fluctuations. First, the $X^i$ have vevs, so that expanding the action in fluctuations is more involved. Secondly, the system is a gauged quantum mechanical system. This means that there are zero modes that should be projected out of the dynamics. Finally, for the displaced fuzzy spheres we are not at an  extremum of the potential, so the gradient of the potential does not vanish. This means that the Hessian that determines quadratic fluctuations is not invariant under non-linear field redefinitions, unlike in the case of an extremum of the potential where the Hessian is a symmetric tensor on the tangent space of the corresponding configuration point. This is potentially problematic for the removal of the zero modes, as the group of  $U(N)$ rotations of the configuration gives us a non-linear geometric space.

All of these problems are solvable in practice. What we need to do is to argue that our fluctuations are orthogonal to linearized gauge transformations on a particular configuration defined by a background. Since we have a metric on the configuration space defined by the kinetic term, this is a well defined procedure. 

We expand the fluctuations of the off-diagonal blocks as follows\footnote{An equivalent way of doing this computation is to expand the fluctuations in the basis of eigenstates of the $b=0$ problem, which was originally solved in \cite{VRetal}. We outline this alternative derivation in the appendix.}
\begin{eqnarray}
X^3&\simeq& L^3 + b\begin{pmatrix}0&0\\
0&1\end{pmatrix} + \sum_{\ell,m} \delta x^3_{\ell m} Y_{\ell m} + (\delta x^3_{\ell m})^* Y^\dagger _{\ell m}\,,\cr
X^+& \simeq& L^+  +\sum_{\ell,m} \delta x^+_{\ell m-1} Y_{\ell m} + (\delta x^-_{\ell m+1})^* Y^\dagger _{\ell m}\,,\cr
X^-&\simeq& L^- + \sum \delta x^-_{\ell m+1} Y_{\ell m} + (\delta x^+_{\ell m-1})^* Y^\dagger _{\ell m}\,.
\label{Xexp}
\end{eqnarray}
Notice that we  have shifted the index $m$ in the coefficients of the $X^\pm$ fluctuations by $\pm 1$. The reason for doing this is that the matrices $L^+$ and $L^-$ usually are associated with angular momentum one. However the configurations with $\delta x=0$ are spherically invariant, so the $L^+$ and $L^-$ matrices should be associated to having no spin: in this way the spin of the matrix is cancelled by the spin of the label.  
The kinetic term for the longitudinal fluctuations will be then given by
\begin{equation}
{\cal L}^{(X)}_{kin}= \frac{1}{2} \sum_{\ell,m} |\delta \dot x^3_{\ell m}|^2 + \frac 12 |\delta \dot x^+_{\ell m-1}|^2+\frac 12 |\delta \dot x^-_{\ell m+1}|^2\,.
\label{eq:kinetic}
\end{equation}
Notice that the metric on these fluctuations is diagonal, but the coefficients are not one. This is important for evaluating frequencies and for projecting out the gauge fluctuations. 

Now let us perform a gauge transformation that is off-diagonal, with parameters $\delta \theta_{\ell m} Y_{\ell m}+ h.c.$.  This is necessary for the generator to be hermitian. We find that 
\begin{equation}
\delta_\theta X^3 = i [\delta \theta_{\ell m} Y_{\ell m}+ h.c.,X^3]=  -i (m-b) \delta \theta_{\ell m} Y_{\ell m}+ i (m-b) \delta\theta_{\ell m}^* Y^\dagger_{\ell m}\,.
\end{equation}
Again, notice that $b$ just shifts the value of $m$ for this calculation. 
Similarly we find that 
\begin{eqnarray}
\delta_\theta X^+ &=& -i \sqrt{(\ell -m)( \ell+m+1) } \delta \theta_{\ell m} Y_{\ell m+1}+ i \sqrt{(\ell + m)( \ell-m+1) } \delta \theta^*_{\ell m}Y^\dagger_{\ell m-1}\,,\cr
\delta_\theta X^-&=& (\delta_\theta X^+)^\dagger \,.
\end{eqnarray}
We then require that the allowed $\delta x^i_{\ell m}$ are orthogonal to the $\delta \theta_{\ell m}$ variations of the configuration. The conjugate variables to these rotations vanish. Classically this means that for fluctuations proportional to $\delta \theta_{\ell m}$ we have to impose the constraint $\dot {\delta \theta}_{\ell m}=0$, so that $\delta \theta_{\ell m}=0$.

These gauge transformations are unphysical and are projected out by the Gauss law constraint. To match our labeling, we should replace the dummy index  $m\to m-1$ or $m\to m+1$ in the various terms in the expansion of $\delta_\theta X^+$ and $\delta_\theta X^-$. We get this way that
\begin{eqnarray}
\delta_\theta X^+ &=& -i \sqrt{( \ell+m)(\ell  -m+1) } \delta \theta_{\ell m-1} Y_{\ell m}+ i \sqrt{( \ell-m)  (\ell  + m+1)} \delta \theta^*_{\ell m+1}Y^\dagger_{\ell m}\,,\cr
\delta_\theta X^-&=& (\delta_\theta X^+)^\dagger 
\end{eqnarray}
and we see the consistency of the conventions of spin labeling, for the $\delta \theta_{\ell m}$ appear in the same way as the $\delta x^i_{\ell m}$ in the expansion of the fields. Notice how in the gauge variations the coefficient of $Y_{\ell \ell}$ in $\delta_\theta X^-$ vanishes, and also the one of $Y_{\ell, -\ell}$ in $\delta_\theta X^+$. This is the object with maximum helicity (total spin along the $3$-axis) at fixed $\ell$. It has helicity $\ell+1$ and $-\ell-1$ respectively. From our previous considerations these are the modes that are most likely to become very light. We will see that they can indeed become tachyonic for some values of $b$.

Now we move on to analyze the potential for these fluctuations. Using the following identities
\begin{eqnarray}
&& i[X^2,X^3]+X^1+i\left(i[X^3,X^1]+X^2\right)=[X^+,X^3]+X^+\,,\cr
&& i[X^2,X^3]+X^1-i\left(i[X^3,X^1]+X^2\right)=-[X^-,X^3]+X^-\,,\cr
&& i [X^1,X^2]= \frac 12 [X^-,X^+]\,,
\end{eqnarray}
the potential can be rewritten as
\begin{equation}
V^{(X)}_{BMN}= \frac 12 \tr\left[   \left( \frac 12 [X^-,X^+]+X^3\right)^2
+ ( [X^+,X^3]+X^+)(- [X^-,X^3] +X^-)\right]\,.
\end{equation}
If we expand in quadratic fluctuations, we can expand each term in the square to linearized order and get that way some quadratic terms. There is an additional term that arises because
when we turn on $b$ the background does no longer satisfy $[X^+,X^-]=2X^3$. The other two equations are satisfied. These contributions from the potential not being at a minimum affect the coefficient of $\delta x^+\, \delta x^-$ in the quadratic terms. 

Writing (\ref{Xexp}) in blocks we have that, schematically, 
\begin{equation}
\delta X^+ = \begin{pmatrix} 0& \delta x^+ Y\\
(\delta x^{-})^*Y^\dagger &0 \end{pmatrix}\,,
\end{equation}
plus a similar expansion for $\delta X^-$. To expand to quadratic order the total potential in off-diagonal modes there are two contributions: those that are linear in the fluctuations in each term of the potential that is squared and those that are quadratic in the fluctuations. The linear terms are off-diagonal, the quadratic terms are block diagonal.
The following intermediate calculations are useful before giving the final answer:
\begin{eqnarray}
&&\delta X^3 + \frac 12[L^-, \delta X^+]- \frac 12 [L^+, \delta X^-] =\nonumber\\
&&\sum_{\ell,m} \left( \delta x_{\ell m}^3 +\frac 12 \sqrt{(\ell-m) (\ell +m+1)}\delta x^+_{\ell m} - \frac 12 \sqrt{(\ell+m)(\ell -m+1)} \delta x^-_{\ell m}\right) Y_{\ell m} + h.c.\,, \cr
&& \cr 
&& [\delta X^+,\vev{X^3} ]+ [L^+, \delta X^3]+\delta X^+=\nonumber\\
&&\sum_{\ell,m}\left((b-m+1) \delta x^+_{\ell ,m-1} + \sqrt{(\ell+m)(\ell-m+1)} \delta x^3_{\ell, m-1} \right )Y_{\ell m} \nonumber \\
&& \hskip 1cm +\left((m-b+1) ( \delta x^-_{\ell m+1})^*- \sqrt{(\ell-m)(\ell+m+1)}(\delta x^3_{\ell m+1})^*\right) Y_{\ell m}^\dagger\,,
\end{eqnarray}
and a similar equation involving $\delta X^-$.
All these terms are off-diagonal (being proportional to the $Y_{\ell m}$). 
Notice that because of our conventions, only the same values of $\ell, m$ appear in all of the coefficients of these linear terms: this is, the mixing of modes only mixes the same values of $\ell, m$. This simplification makes the problem very tractable for these modes as well. In the end, we need to understand how three modes mix, but one such mode is projected out because of the gauge constraint. The general mass reduces to diagonalizing a $2\times 2$ matrix for each value of $\ell,m$

This can be combined to give  a block diagonal term in $\frac12[ X^-,X^+]+ X^3$ proportional to 
\begin{equation}
\begin{pmatrix}
0&0\\
0& b
\end{pmatrix}+\frac 12 \begin{pmatrix} \delta x^-  (\delta x^-)^*YY^\dagger-\delta x^+ (\delta x^+)^*  Y  Y^\dagger &0\\
0 &\delta x^+ (\delta x^+)^*  Y^\dagger  Y- \delta x^-  (\delta x^-)^*Y^\dagger Y
\end{pmatrix}\,.
\end{equation}
When we square and take traces, expanding to quadratic order in fluctuations,  this gives us a contribution to the mass matrix equal to
\begin{equation}
\frac 12 b ( \delta x^+ (\delta x^+)^*- \delta x^-  (\delta x^-)^*)\,.
\label{bcontr}
\end{equation}
Notice that the contribution from this term is negative or positive for different modes depending on the sign of $b$.

The kinetic term \eqref{eq:kinetic} suggest that we normalize the fields slightly differently, $\delta x^{\pm} = \sqrt 2\, \delta X^{\pm}$, to have canonical normalizations for every mode. The terms in the expansion can be rewritten as
\begin{eqnarray}
\delta x_{\ell m}^3 +\frac 12 \sqrt{(\ell-m) (\ell +m+1)}\delta x^+_{\ell m} - \frac 12 \sqrt{(\ell+m)(\ell -m+1)} \delta x^-_{\ell m}\nonumber \\
= \begin{pmatrix}  1\,, ~~ &  \sqrt{\frac{(\ell-m)(\ell +m+1)} { 2}} \,,~~ & 
 -\sqrt{\frac {(\ell+m) (\ell+1 -m)}{ 2}}\end{pmatrix}\begin{pmatrix}
 \delta x^3_{\ell m}\\
\delta X^+_{\ell m}\\
\delta X^-_{\ell m}
 \end{pmatrix}\equiv  V^3 \, \delta X\,.
\end{eqnarray}
Similarly we find a $V^+$ and $V^-$ given by
\begin{eqnarray}
V^+ &=&\begin{pmatrix}\sqrt{(\ell-m)(\ell +m+1)}\,, ~~ & {\sqrt 2} {(b-m)}\,, ~~ & 0\end{pmatrix}\,,\cr 
V^- &=& \begin{pmatrix}- \sqrt{(\ell+m)(\ell -m+1)} \,,~~ &  0\,, ~~ &{\sqrt 2}{(m-b)}\end{pmatrix}\,.
\end{eqnarray}
We have shifted $m\to m\pm 1$ in the equations above so that we are comparing the same coefficients of $\ell, m$.

The mass matrix is given by squaring these vectors and adding them together including also the contribution of (\ref{bcontr})
\begin{equation}
\omega^2_{\ell m} =  (V^3)^\dagger V^3 + \frac 12 ( V^+)^\dagger V^++\frac 12(V^-)^\dagger V^-+ \begin{pmatrix} 0&0&0\\
0&b&0\\
0&0&-b
\end{pmatrix}\,.
\end{equation}
The end result is given by
\begin{equation}
\omega^2_{\ell m} = 
\begin{pmatrix}
1+\ell+\ell^2-m^2&\ & (b-m+1)\Lambda_-&\ & (b-m-1)\Lambda_+\\
(b-m+1)\Lambda_-&& b+(b-m)^2+\Lambda_-^2&& -\Lambda_+\Lambda_-\\
(b-m-1)\Lambda_+&&  -\Lambda_+\Lambda_-&&  -b+(b-m)^2+\Lambda^2_+
\end{pmatrix}\,,
\end{equation}
where we have defined the shorthands
\begin{equation}
\Lambda_\pm\equiv \sqrt{\frac{(\ell\pm m)(\ell \mp m+1)}{2}}\,.
\end{equation}
Of particular interest to us is when $m=\ell +1$  for $\delta X^-_{\ell, m}$ ({\it i.e.} $\Lambda_+=0$), and when $m=-\ell -1$ for $\delta X^+_{\ell, m}$ ({\it i.e.} $\Lambda_-=0$). For these cases there is no mixing with any other mode and these fields have maximum spin in the $3$-direction for fixed $\ell$. We have already argued why these modes are important. Their masses are given by
\begin{eqnarray}
&& (\omega^-_{\ell, \ell+1})^2= -b +(b-\ell-1)^2\,,\cr 
&& (\omega^+_{\ell, -\ell-1})^2= b +(b+\ell+1)^2\,.
\label{spectrum}
\end{eqnarray}
These modes are tachyonic for $b= \pm(\ell +1) $ on an interval for $b$ of order $\sqrt \ell$. Notice that there is a tower of tachyonic modes for each $b$ labeled by $\ell$ with a quadratic 
dispersion relation. This can be interpreted as a tower of tachyonic  modes on a circle in the presence of some holonomy for a gauge field under which these fields are charged. Other modes for which $m$ is not maximal in the sense above are not tachyonic.

From the equation for the masses above we still need to project out the gauge variations. This is straightforward, but tedious.  If we call the projection matrix that projects onto the gauge degrees of freedom as $\Sigma_{\ell m}$, then $1-\Sigma_{\ell m}$ is the projection in the orthogonal components. The mass matrix we need is then given by
\begin{equation}
\omega^2_{\ell m,  \, \tiny{phys}}= (1-\Sigma_{\ell m})\omega^2_{\ell m}(1-\Sigma_{\ell m}) 
\end{equation}
The precise expressions are not very illuminating. However, none of the modes that appear this way are tachyonic, except the ones that we have already discussed.

We can also check that the eigenvalues of the above matrix are $\ell^2, (\ell+1)^2, 0$ when we set $b=0$ (as originally found in \cite{VRetal}) as a consistency check. For $b=0$ the modes with zero eigenvalue are the gauge zero modes. For $b\neq 0$ these modes seem to become massive (the determinant is not zero), but as we have argued already, this is an artifact of the linearization. After all, expanding to second order in these gauge variations  we find that 
\begin{equation}
 V^{(X)}_{BMN} (b, \delta \theta) \simeq V^{(X)}_{BMN}(b) + \partial_b V^{(X)}_{BMN}(b)(\delta \theta^2) + \ldots\,,
\end{equation}
and the second term only vanishes for $b=0$. However, the potential is invariant, so $b$ must be corrected to second order in gauge fluctuations. This is a non-linear change of variables. This is why it is better to project on directions orthogonal to the gauge transformations than trying to sort this second order variation and how it affects the metric of the other modes.

It is clear that when we consider the above result, we should organize the modes according to the following criteria. If the two fuzzy spheres intersect for some $b >0$, we should fix $K= \ell -m$ for the modes where $b\simeq m$ (those that are near the intersection). For each such $K$ we get a tower of states on a circle labeled by the different values $\ell$ (or $m$) (there are two states for general $\ell, m$). These look like a tower of fields in one dimension. For $b<0$ we would do the same by  fixing $K=\ell +m$ for the modes near the intersecting fuzzy spheres.

Notice that this result is very similar to results that have been  found in other matrix models with fuzzy spheres \cite{Azuma:2007jr}. Since in their case the solutions with displaced fuzzy spheres are critical points of the potential, the issues with zero modes do not appear.


\subsection{Interpretation from ${\cal N}=4$ super Yang-Mills}

We would like to look at what we have obtained so far from the point of view of ${\cal N}=4 $ super Yang-Mills on $S^3\times \mathbb{R}$. The results above are expected to give details of an $SU(2)_L$ invariant subsector of this theory on the sphere. After all, the BMN dynamics is exactly the truncation to the $SU(2)_L$ invariant subsector of the field theory. Thus, any dynamical feature present in the $SU(2)_L$ invariant truncation is a feature of the full ${\cal N}=4$ super Yang-Mills on the sphere. In particular, the unstable modes  
we found in the matrix model are necessarily unstable modes in the full field theory.

For all the fluctuations we have studied, notice that, apart from the bounds on the total magnetic moment, $|j_1\pm j_2|$, our results are essentially independent of the values $n_1$ and $n_2$. If we think back at how the plane wave matrix model is related to a truncation of ${\cal N}=4$ super Yang-Mills on  $S^3$ we realize that we should not be surprised by this fact. The splitting into $n_1$ and $ n_2$ as fuzzy spheres is artificial in ${\cal N}=4 $ super Yang-Mills. The fuzzy spheres are gauge transformations of the vacuum. However $b$ matters: its presence gives states with non-zero energy. 
In the  field theory setup the total angular momentum of physical states also matters, but not the details of  the splitting into $n_1, n_2$, as these would just give a multiplicity of components with different masses from different angular momentum objects. 

To make things precise, we should remember that the fields $X^i$ arise from the gauge connection on $S^3$, while the fields $Y^a$ arise from the constant scalar modes of the 
field theory. The way to do this is simple: since we have an $S^3$ sphere, the spatial components of the gauge connection can be written in a basis of left invariant one-forms $e^i$
\begin{equation}
{\cal A}(t) = X^i(x,t) e^i\,.
\end{equation}
The spatial component of the field strength (chromo-magnetic field) is given by
\begin{equation}
d{\cal A}+{\cal A}\wedge {\cal A}= dX^i \wedge e^i + X^i de^i+ \frac 12 [X^i, X^j] e^i \wedge e^j\,.
\end{equation}
Restricting to $SU(2)$ invariant states requires that $X^i$ be constant. Thus on such configurations the first term vanishes. However, the one-forms are not exact $de^i\neq 0$. Instead they satisfy the Maurer-Cartan equations. Thus the second and third terms do not vanish. Even for constant  diagonal $X^i$ there is a non-zero result for the magnetic field. Considering the magnetic field squared part of the Hamiltonian gives an expression that is exactly the term in the potential given by equation \eqref{eq:VXBMN}, after one accounts for various normalization issues. 
Thus, a non-trivial diagonal displacement in the $X^i$ (of size $b$) corresponds to a chromo-magnetic field of strength proportional to $b$ (the commutator squared term vanishes for this configuration).

Notice that in such a description of the theory the parameter $b$ would correspond to a magnetic field that leaves invariant an $SU(2)_L$ of rotations on the $S^3$ (the electric field is related to the time derivative of the $X^i$). It also leaves invariant an $U(1)_R$ of rotations along the direction of said magnetic field. For such a system the $SU(2)$ spherical harmonics under $SU(2)_L$ and the $U(1)$ quantum numbers completely specify the spectrum for the scalar fluctuations, after all, these are originally classified as the $(n,n)$ representations under $SU(2)\times SU(2)$. The first quantum number determines $n$, while the second quantum number denotes the $z$-component of spin in the $SU(2)_R$. Moreover, this symmetry  splits the degeneracy in energies for fluctuations with different $U(1)_R$ completely. 

When truncating to $SU(2)_L$  invariant states the choices of fuzzy spheres matter again, but this does not affect the energies, just the  $SU(2)_L$ labels of the fields that get twisted.  As we also argued, the $SU(2)_R$ labels are tied very closely to the $SU(2)_L$ labels of the field around the trivial vacuum.  Since the system preserves the $SU(2)_L$ symmetry, we find that the splitting of modes into $SU(2)_R$ representations is preserved in the trivial vacuum. This is not obvious in the matrix model computation since we were turning on a vacuum expectation value that breaks the $SU(2)_R$ symmetry to $U(1)_R$ and in principle could induce mixing between modes with different $SU(2)_R$ representations. This non-mixing between modes looks like a happy coincidence from the matrix quantum mechanics point of view. Here we see that there is a symmetry reason from super Yang-Mills for that to happen. Moreover, each such $(n,n)$ can contribute at most one singlet under $SU(2)_L$ for each value of the $U(1)$ angular momentum.

We should also ask what kind of instability do the modes between intersecting fuzzy spheres represent in super Yang-Mills. Clearly, the fuzzy sphere is a gauge artifact, but not the presence of a magnetic field. So we should ask what kind of instabilities can arise in the presence of a constant chromo-magnetic field on a sphere. If we take $b$ large, this corresponds to a large magnetic field. By thinking in terms of engineering units, the magnetic field $b$ generates a scale in the system that,  if $b$ is large, is much shorter in length than the radius of the sphere. Under such conditions we can ignore the radius of the sphere and treat the magnetic field as if it were constant in space.  The only scale in the problem is associated to $b$. The mass squared of the unstable mode should therefore be linear in $b$ (just a result of dimensional analysis). 
If one has momentum $p$ along the direction of the magnetic field, the spectrum gives modes with frequencies of the form $\omega_{\ell m}^2 = -|b|+ p^2$. This is exactly the result we found in equation \eqref{spectrum}, if we identify $p= \ell+1\pm b$.

 Such a result is well known. It is the Nielsen-Olesen instability \cite{Nielsen:1978rm} (see also \cite{Leutwyler:1980ma}). This instability is due to the fact that gluons charged under the field that acquired a magnetic field have a large magnetic moment that is enough to drive them tachyonic. For scalars there is no magnetic moment, and the localization effects in a magnetic field cost energy via the uncertainty principle, so these modes are not tachyonic, just as we found for the fluctuations of the $Y^a$ fields. For fermions, one can get zero modes, which is the familiar connection between massless fermions and index theory. Incidentally, it is well known that it is the magnetic moment contribution to the $\beta$-function of non-abelian gauge theories that turns the sign contribution for vector particles relative to scalar particles. It was argued that this would also lead to an effective action where the chromo-magnetic field condenses \cite{Savvidy:1977as}. The Nielsen-Olesen instability would destroy that type of order.


\section{Time dependence for the tachyonic modes}
\label{sec-5}

So far in our analysis we have considered configurations that are frozen in time. We have seen that tachyons can form in the region where the two fuzzy spheres intersect and we have argued that they do not mix with any other mode. At this point an interesting question to ask is what happens to these tachyons as we let our system evolve. In this section we present a straightforward analysis for the simplest possible trajectory, namely an oscillation of the spheres along one fixed direction. 

Consider two fuzzy spheres of sizes $j_1$ and $ j_2$. The tachyons appear for any integer value from $|j_1-j_2|$ all the way to $j_1+j_2$ if the smaller sphere is allowed to oscillate from the center  all the way to the outside of the larger sphere.  As we have computed in (\ref{spectrum}), the mass for these tachyons is given by
\begin{equation}
(m_\ell^\pm(t))^2 = (\ell+1\pm b(t))^2\pm b(t)\,,
\end{equation}
where we do not write the label $m$, which is set equal to $\mp(\ell+1)$. We take $b(t)= \tilde b \sin t$, so to have a simple sinusoidal motion along the $X^3$ direction. The motion in $t$ is periodic with period $2\pi$, so the full analysis can be restricted to the interval $t\in \{0, 2\pi\}$. We need to solve the differential equation
\begin{equation}
\ddot q_\ell (t) + (m^\pm_\ell(t))^2 q(t)=0\,.
\label{mathieu}
\end{equation}
This is the equation for a harmonic oscillator with time dependent mass. It is somewhat similar\footnote{Albeit more complicated and not generically solvable in terms of elementary  functions.} to the Mathieu equation that describes parametric resonances and, in cosmology, the fluctuations of the inflaton around the minimum of the potential during preheating \cite{Kofman:1994rk}.

In general there are two linearly independent solutions to the equation (\ref{mathieu}), which we call $q_1(t)$ and $q_2(t)$.  We can relate the initial time problem (at time $t$) to the problem at one period later (at time $t+2\pi$) using a periodicity matrix
\begin{equation}
\begin{pmatrix} q_1(t+2\pi)\\
q_2(t+2\pi) 
\end{pmatrix}= \begin{pmatrix} A&B\\
C&D
\end{pmatrix} \begin{pmatrix} q_1(t)\\
q_2(t) 
\end{pmatrix}\,.
\end{equation}
This equation can be diagonalized, so we can choose the solutions to be eigenvalues of the matrix above. The Wronskian of the solution is constant, so the matrix transforming between one and the other has determinant equal to one. Also, since the differential equation has real coefficients, the solutions can be made real and in that case the matrix above is real as well. Hence, the eigenvalues are either real or unitary. These eigenvalues serve as Lyapunov exponents for the classical periodic orbit. When the eigenvalues are unitary the system is stable, when the eigenvalues are real the system is unstable. 

The system can be interpreted also as a Schr\"odinger problem with fixed energy in a periodic potential,\footnote{For an elementary treatment see \cite{kittel}.} which is the negative of the $(m^\pm_{\ell}(t))^2$ function. If the solutions are quasi-periodic (the eigenvalues are in the unit circle), one of the functions is identified with positive frequency and the other one is identified with negative frequency modes. This is the case where the functions $q_\ell(t)$ belong to a band of the periodic potential. 

Generically, if there are regions where the mode is tachyonic, the corresponding Schr\"odinger particle needs to tunnel through the barrier. This phenomenon generically leads to the property that the eigenvalues of the matrix above are non-unitary, with the tunneling amplitude characterizing the growth of the signal. We can in general estimate this using a WKB approximation. 

 The large eigenvalue tells us how the modes grow around these periodic solutions and it describes the discrete time dependence of the instability under various oscillations. The two linearly independent solutions can also be thought of as coefficients of raising/lowering operators. The matrix computed in this basis is a Bogolubov transformation for each period and the amplitude growth correlates with the amount of particle creation between oscillations. 

\begin{figure}[ht]
\begin{center}
\includegraphics[scale=1]{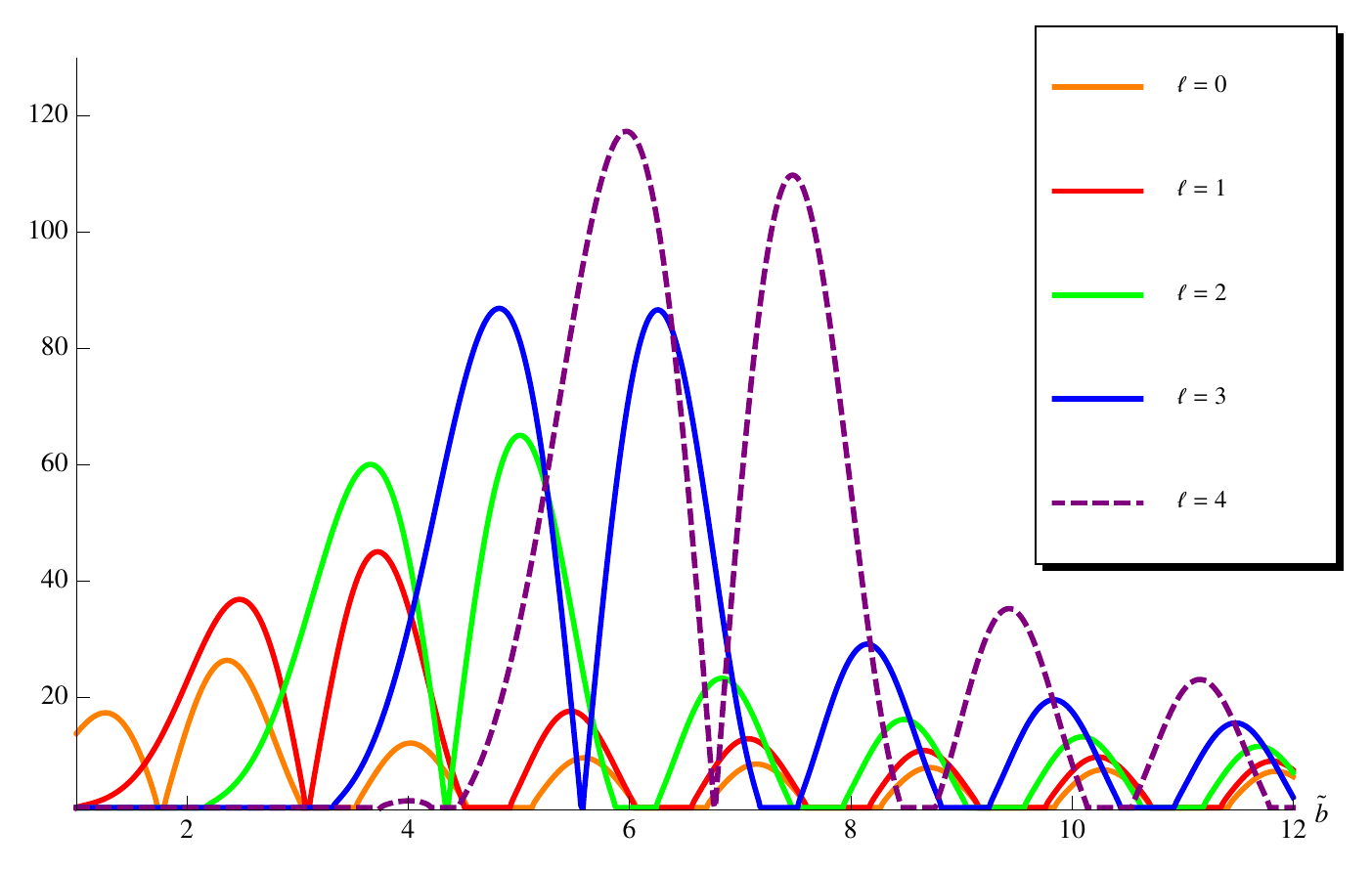}
\caption{ Here we show the norm of the maximum eigenvalue of the periodicity matrix for various values of $\ell$ as a function of the amplitude $\tilde b$, labeling the horizontal axis. We clearly see the band structure of the problem and that 
the largest value of $\ell$ is typically the one with most amplification so long as it is tachyonic. 
\label{fig:tachyonamp}}
\end{center}
\end{figure}

\begin{figure}[ht]
\begin{center}
\includegraphics[scale=0.95]{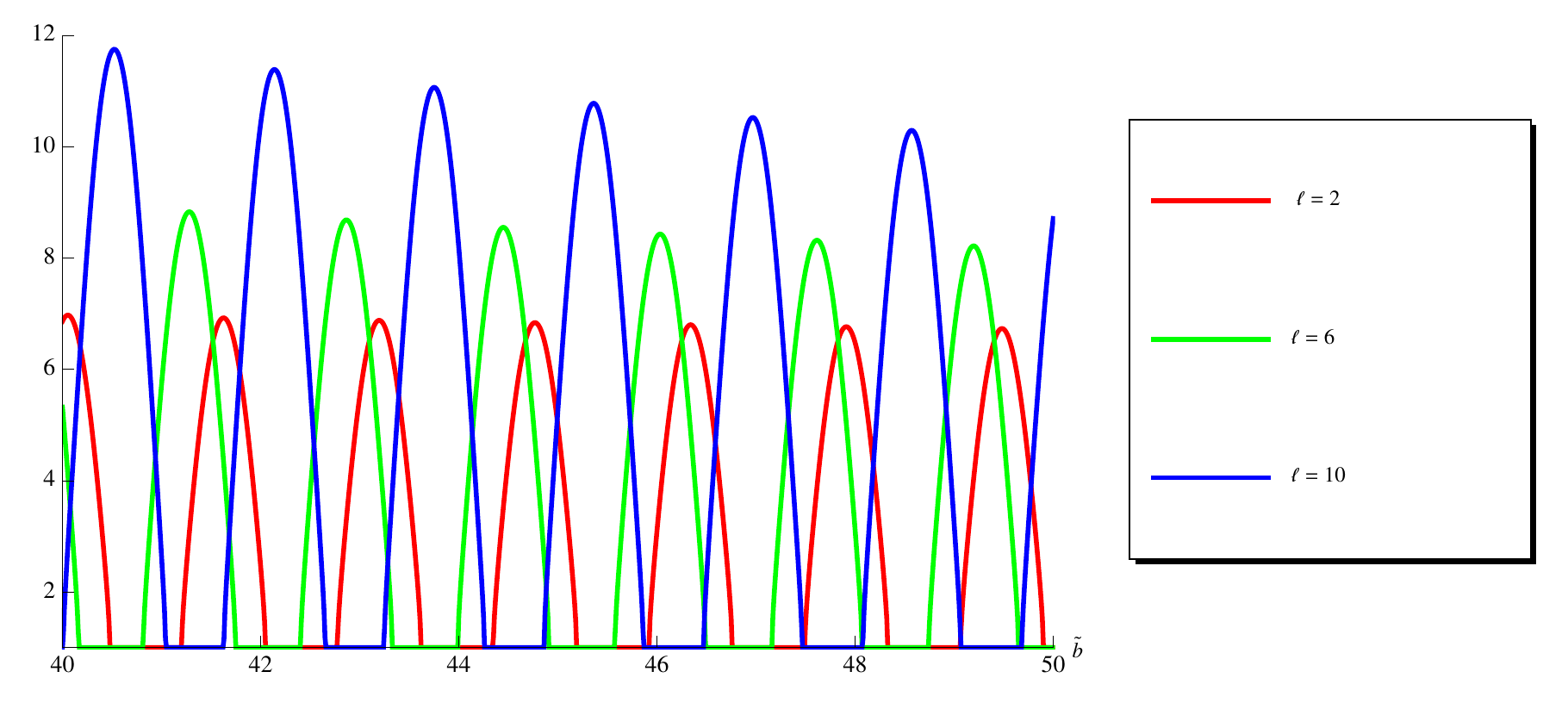}
\caption{ Here we show the maximum eigenvalue of the periodicity matrix for various values of $\ell$, where we explore the problem in the region of large amplitude $\tilde b$, labeling again the horizontal axis. Asymptotically the amplification becomes of order one, since the time during which each mode is tachyonic on a single oscillation becomes very small.  
\label{fig:tachyonamplong}}
\end{center}
\end{figure}

For us, the most important question to ask is which of all the modes above grows the fastest, as these modes will dominate the initial stages of the brane collapse problem. It is not hard to figure out that the tachyon with the highest $\ell$ will generally dominate. First, it will be tachyonic for the longest time during the periodic trajectory, not only because the range of $b$ where it is tachyonic is larger (recall that this range is of order $\sqrt{\ell}$), but also because the motion of the oscillations is slower at larger displacement. This means that the main condensation of modes will happen between the north pole of one fuzzy sphere and the south pole of the other one. We can do this numerically for various values of the amplitude of oscillation $\tilde b$ and for different values of $\ell$.
This is shown in Figures \ref{fig:tachyonamp} and \ref{fig:tachyonamplong}.

Notice also that because condensation happens for various modes with different $\ell$, the system classically breaks the rotation symmetry around the axis of symmetry for small perturbations on the off-diagonal modes. Each of these is expected to decohere from the others once the non-linearities set in, because the classical trajectories diverge rapidly from each other. Also, as seen in the figures, any small amount of back-reaction can move modes between the oscillation bands and the growth bands. Let us assume that this will happen for some fixed value of the off-diagonal perturbation in the classical picture. The time to reach this value will depend on the initial amplitude of the off-diagonal mode. If we begin in an adiabatic ground state for these modes  in the non-tachyonic region, we can estimate the initial size of fluctuations by a harmonic oscillator wave function. This is a power-law function of $\hbar$.
The growth is exponential (and can easily be of a few orders of magnitude per oscillation), so the time growth for quantum perturbations around a fixed classical periodic orbit to the stages where back-reaction becomes important is logarithmic in the $\hbar\to 0$ limit. As we increase $\hbar$, the time to reach back-reaction is shorter. In the very quantum regime the approximations used here break down and a more robust formalism with quantum back-reaction needs to be developed (an initial attempt  of such a setup has been recently pursued in \cite{Asplund:2010xh}).  This suggests that in the strong coupling regime thermalization could be very fast. 


 \section{Conclusion}
\label{sec-6}

The purpose of this paper was to initiate a study of the process of formation and thermalization of black holes, using the gauge/gravity correspondence. Since these are notoriously complicated problems to address in ordinary field theories, our strategy has been to focus our attention on simplified models with a reduced number of degrees of freedom. 

The prototypical example of such models is the BFSS matrix quantum mechanics describing M-theory on the discrete light-cone quantization of flat space. Unfortunately, this model possesses some features that make its study extremely challenging. These features include the presence of flat directions in the potential (which give rise to a continuous spectrum and to the difficulty of distinguishing between single-particle and multi-particle states) and the absence of a tunable coupling constant. Moreover, the wave function of bound states in this model is not known, making it impossible to describe scattering processes in the regime of high energy and small impact parameters where the non-linearities of gravity become important and black hole formation takes place.\footnote{Most of the work in the past aimed at matching the loop expansion of the BFSS matrix model with perturbative expansions in gravity has been mostly limited to linearized examples or to terms protected by non-renormalization theorems, see for example \cite{Becker:1997xw} and \cite{wati} for a review.} 

We have therefore considered a different model, the BMN matrix model given by (\ref{BMNaction}), where many of the problems that plague the BFSS model no longer occur. For example, the mass terms in the BMN model lift flat directions and give rise to a discrete spectrum with an isolated set of classical vacua, the fuzzy spheres that have been the central ingredient of our setup. In the limit of large mass, these different vacua are divided into superselection sectors described by harmonic oscillators and the spectrum of their fluctuations can be studied. While the supersymmetric vacua of the model consists of concentric spheres, one can also consider non-BPS configurations obtained by displacing the centers of the spheres. These are obtained by turning on the diagonal modes that control the center of mass motion of the spheres without deforming their shape. In this way one can setup a scattering problem, that might be used as a proxy, under computational control, for the high energy scattering of particles in gravity. 

We have started our analysis by considering two fuzzy spheres, displaced along a direction in their world-volume and frozen in time in that position. The formation and thermalization of a black hole can be associated to copious particle production in off-diagonal modes of a configuration.
The presence of classical tachyons makes the analysis simpler as we get a classical instability that can in principle drive the system towards thermalization.   Expanding the fields in fuzzy spherical harmonics, we have found that the modes with the maximal angular momentum along the direction of the displacement become tachyonic at the intersection locus between the two spheres. Interestingly, this instability has a four-dimensional interpretation. It can be regarded in fact as a Nielsen-Olesen instability in ${\cal N}=4$ super Yang-Mills, of which the BMN model is a truncation. The role of the background magnetic field is played in our system by the displacement vector. 

An obvious generalization of our initial conditions would be to allow for a displacement of the fuzzy spheres also along the transverse directions (this corresponds to turning on some $\vev{Y^a}\neq 0$). The analysis in this case is slightly complicated by the fact that the fluctuations of the $X^i$ and the $Y^a$ fields get coupled and the diagonalization problem becomes less straightforward. One could also try to include the fermionic modes, which have not been considered here. 

In the last part of the paper, we have allowed our initial configuration to evolve in time describing a periodic motion, with the two fuzzy spheres oscillating along the direction of the displacement and crossing each other repeatedly. This is the simplest trajectory to study and gives rise to a dynamics that is somewhat similar to the physics of preheating during inflation. We have argued that the tachyonic modes that form at the intersection locus between the spheres typically get reinforced after each period of the oscillation. The  growth is exponential, and can potentially give rise to a fast thermalization of the other degrees of freedom living on the spheres. So if these instabilities would cause the BMN model to thermalize fast, it is suggestive that the Nielsen-Olesen instability could drive fast thermalization in QCD processes like heavy-ion collisions. This  thermalization is observed in experiments at RHIC  \cite{Ackermann:2000tr} and it is argued to happen in a very short time scale \cite{Heinz:2004pj}.

A better understanding of the details of this dynamics is surely desirable and this paper can be considered as a first step toward this ambitious goal. In particular, we find it extremely interesting to try to estimate from our setup the time scales characterizing the various phases of the black hole evolution, for example the thermalization time. To this regard, it has recently been conjectured \cite{SS} (prompted by the analysis in \cite{preskill}) that black holes are the `fastest scramblers' in nature. By fastest is intended  that the thermalization time scale is logarithmic in the number of degrees of freedom of the system, rather than a  power-law, as was originally proposed in \cite{page}. Our hope is that it might be possible to check this claim using the ideas and techniques we have presented in this paper.

An even more involved scenario would start with a configuration that carries angular momentum on the plane. Such a scenario might lead to different black hole shapes. The time dependent analysis we performed would be more complicated because there is more mixing between modes (the system does not preserve an azimuthal symmetry). 

We conclude by repeating the observation in the Introduction that the BMN matrix model is amenable to being put on a computer. We can think of implementing numerical simulations of our system, where we define some initial configuration of displaced fuzzy spheres and let them evolve. Such simulations should provide us with a more accurate description of the details of the dynamics that follows the formation of the tachyons and shed light on what happens during the first phases of the evolution of the black holes.  We are currently looking into this.


\subsection*{Acknowledgements}
D.B. would like to thank C. Asplund,  D. Kabat, J. Maldacena, E. Silverstein, and H. Verlinde for various discussions related to this work. D.T. is grateful to S. Giddings for several instructive discussions on high energy scattering in gravity and to N. Iizuka for collaboration on related topics. D.B. would like to thank the Simons Center for Geometry and Physics, where some of this work was carried out. Work supported in part by DE-FG02-91ER40618. D.T. was also supported by PHY04-56556 and DE-FG02-95ER40896. 

\appendix
\section{An alternative derivation of the longitudinal spectrum}

In this appendix we present an alternative derivation of the spectrum of longitudinal fluctuations (\ref{spectrum}). The idea is to expand the fluctuations in the basis of eigenstates of the $b=0$ system. 

We start by rewriting the quadratic potential for the $\delta X^i$ fluctuations as
\bea
V_{BMN}^{(X)} 
= \frac{1}{2}\tr\left(\left(\delta X^i+i \epsilon^{ijk}\left[\vev{X^j},\delta X^k\right]\right)^2
-i\,  \epsilon^{ij3}\delta X^i\left[\begin{pmatrix} 0&0 \\ 0 & b\end{pmatrix},\delta X^j\right]\right)
 \,.
 \label{pot_so3}
\eea
The case with $b=0$ was worked out in detail in \cite{VRetal}, where it was found that the solution to the eigenvalue problem
\be
\delta X^i_{(\alpha)}+i \epsilon^{ijk}\left[L^j,\delta X^k_{(\alpha)}\right]=\lambda_{\alpha} \, \delta X^i_{(\alpha)}
\ee
is $\lambda_{\alpha}=(0,-\ell,\ell+1)$ with
\bea
&& \delta X^+_{(0)}=\gamma_{\ell m}\sqrt{\frac{(\ell-m)(\ell+m+1)}{\ell(\ell+1)}}\,Y_{\ell m+1} + h.c.\,,\cr 
&& \delta X^-_{(0)}=\gamma_{\ell m}\sqrt{\frac{(\ell+m)(\ell-m+1)}{\ell(\ell+1)}}\,Y_{\ell m-1}+ h.c.\,,\cr 
&& \delta X^3_{(0)}=\gamma_{\ell m}\, \frac{m}{\sqrt{\ell(\ell+1)}}\, Y_{\ell,m}+ h.c.\,, 
\cr
&& \cr
&&  \delta X^+_{(-\ell)}=-\alpha_{\ell-1,m}\sqrt{\frac{(\ell+m)(\ell+m+1)}{\ell(2\ell+1)}}\,Y_{\ell m+1}+ h.c.\,,\cr 
&&  \delta X^-_{(-\ell)}=\alpha_{\ell-1,m}\sqrt{\frac{(\ell-m)(\ell-m+1)}{\ell(2\ell+1)}}\,Y_{\ell m-1}+ h.c.\,,\cr 
&&  \delta X^3_{(-\ell)}=\alpha_{\ell-1,m}\sqrt{\frac{(\ell+m)(\ell-m)}{\ell(2\ell+1)}}\, Y_{\ell m}+ h.c.\,, 
\cr
&& \cr
&&  \delta X^+_{(\ell+1)}=\beta_{\ell+1,m}\sqrt{\frac{(\ell-m)(\ell-m+1)}{(\ell+1)(2\ell+1)}}\,Y_{\ell m+1}+ h.c.\,,\cr 
&&  \delta X^-_{(\ell+1)}=-\beta_{\ell+1,m}\sqrt{\frac{(\ell+m)(\ell+m+1)}{(\ell+1)(2\ell+1)}}\,Y_{\ell m-1}+ h.c.\,,\cr 
&&  \delta X^3_{(\ell+1)}=\beta_{\ell+1,m}\sqrt{\frac{(\ell+m+1)(\ell-m+1)}{(\ell+1)(2\ell+1)}}\, Y_{\ell m}+ h.c.\,.
\label{eig3}
\eea
Expanding the off-diagonal fluctuations in the basis of eigenstates above
\be
 \delta X^i = \sum_\alpha  \delta X^i_{(\alpha)}\,,
\ee
the potential (\ref{pot_so3}) becomes
\bea
&& \frac{1}{2}\tr\left(\left(\sum_\alpha \lambda_\alpha  \delta X^i_{(\alpha)}-i \epsilon^{ij3}\left[\begin{pmatrix} 0&0 \\ 0 & b\end{pmatrix}, \delta X^j_{(\alpha)}\right]\right)^2-i\,\epsilon^{ij3}\sum_{\alpha,\beta} \delta X^{i}_{(\alpha)}\left[\begin{pmatrix} 0&0 \\ 0 & b\end{pmatrix}, \delta X^j_{(\beta)}\right]\right)
 \,.\cr &&
\eea
We can write the blocks explicitly
\bea
 &&\hskip -1cm \frac{1}{4}\tr \sum_{\alpha,\beta}\left(
 \lambda_\alpha\lambda_\beta\left( (\delta X^{+}_{(\alpha)})^\dagger\delta X^+_{(\beta)}+ (\delta X^{-}_{(\alpha)})^\dagger\delta X^-_{(\beta)}+ 2(\delta X^{3}_{(\alpha)})^\dagger \delta X^3_{(\beta)}\right) \right.
 \cr && 
\left. 
+ b \left(b+\lambda_\alpha+\lambda_\beta+1 \right) (\delta X^{+}_{(\alpha)})^\dagger \delta X^+_{(\beta)}
+ b \left(b-\lambda_\alpha-\lambda_\beta-1 \right) (\delta X^{-}_{(\alpha)})^\dagger\delta X^-_{(\beta)}
  \right).
\eea
Taking the trace, it is easy to see that the first line of the expression above reproduces the $b=0$ spectrum
\be
\frac{1}{2}\sum_{\ell,m}\ell^2|\alpha_{\ell-1,m}|^2+(\ell+1)^2|\beta_{\ell+1,m}|^2\,,
\label{1stline}
\ee
with the $\gamma_{\ell m}$ modes being the gauge variation zero modes.
The second line gives instead (here we call $\delta x_{(\alpha)}$ whatever is multiplying the spherical harmonics in (\ref{eig3}), including the square roots)
\bea
&& \frac{b}{4} \sum_{\alpha,\beta}\sum_{\ell,m}\left( \left(b+\lambda_\alpha+\lambda_\beta+1 \right) (\delta x^{+}_{(\alpha)})^*_{\ell m} (\delta x^{+}_{(\beta)})_{\ell m}
\right.\cr &&\hskip 4cm \left.
+ \left(b-\lambda_\alpha-\lambda_\beta-1 \right)(\delta x^{-}_{(\alpha)})^*_{\ell m}(\delta x^{-}_{(\beta)})_{\ell m}
\right)\,. 
\label{2ndline}
\eea
Expanding the sums in $\alpha$ and $\beta$ in (\ref{2ndline}) and combining this with (\ref{1stline}), it is straightforward to check that the resulting mass matrix admits the same eigenvalues obtained in (\ref{spectrum}).


\end{document}